\documentclass[9pt,twocolumn,twoside]{opticajnl}
\journal{opticajournal} 

\setboolean{shortarticle}{false}

\usepackage{siunitx}
\newcommand{\SIadj}[2]{\SI[number-unit-product={\text{-}}]{#1}{#2}}
\usepackage{xcolor}

\title{Optimizing broadband microwave absorbers for applications in the 70--200 GHz range}

\author[1]{Gaganpreet Singh}
\author[2]{Rustam Balafendiev}
 \author[4]{Mahesh Singh Bist}
\author[2]{Thomas J.L.J. Gascard}
\author[1]{Gagandeep Kaur}
\author[3]{Vid Primožič}
\author[1, 2,*]{Jon E. Gudmundsson}

\affil[1]{The Oskar Klein Centre, Department of Physics, Stockholm University, SE-106 91 Stockholm, Sweden}
\affil[2]{Science Institute, University of Iceland, 107 Reykjavik, Iceland}
\affil[3]{Faculty of Mathematics and Physics, University of Ljubljana, Ljubljana, Slovenia}
\affil[4]{Department of Electrical Engineering, Eindhoven University of Technology (TU/e), The Netherlands}

\affil[*]{jegudmunds@hi.is}

\begin{abstract}
We present results of an extensive suite of numerical simulations that probe square-tiled microwave absorber performance as a function of material properties, frequency, geometry, and unit cell size. The work, which probes both specular reflection and total absorption, highlights the critical importance of the absorber scale size relative to the incidence wavelength while suggesting that material properties have a comparatively weaker impact on overall performance. We show that some absorber designs can achieve 99.5--99.9\% frequency-averaged absorption across the 70 to 200 GHz range for normal incidence and that low specular reflectance does not necessarily guarantee optimal absorption performance. Our results indicate that exponential, Klopfenstein, and linear impedance tapers provide comparable performance as long as a unit cell size of 1 to 4 mm is chosen. Simulation results are validated against measurements of specular reflectance.
\end{abstract}

\setboolean{displaycopyright}{false} 

\begin{document}
\maketitle

\section{Introduction}
\label{sec:intro}

Broadband microwave absorbers that perform well at cryogenic temperatures have been used in experiments observing the cosmic microwave background (CMB) for quite some time \cite{Hemmati1985, Halpern1986, Lamb1997}. In recent years, several new absorber candidates have been developed \cite{lonnqvist2006monostatic, Wollack2008, wollack2014cryogenic, Chuss2017} and many of those rely on 3D printing or injection molding techniques \cite{petroff20193d, otsuka2021material, xu2021simons, Singh2024, Occhiuzzi2024}. The development of these absorber candidates offers challenges in material design, simulation, and measurement that are met at varying level of detail in the literature. A particularly challenging aspect involves modeling the total absorptance of lossy dielectrics with unit cell sizes that are comparable to the incident wavelength. 

\begin{figure}[b!]
    \centering
    \includegraphics[width=1.0\linewidth]{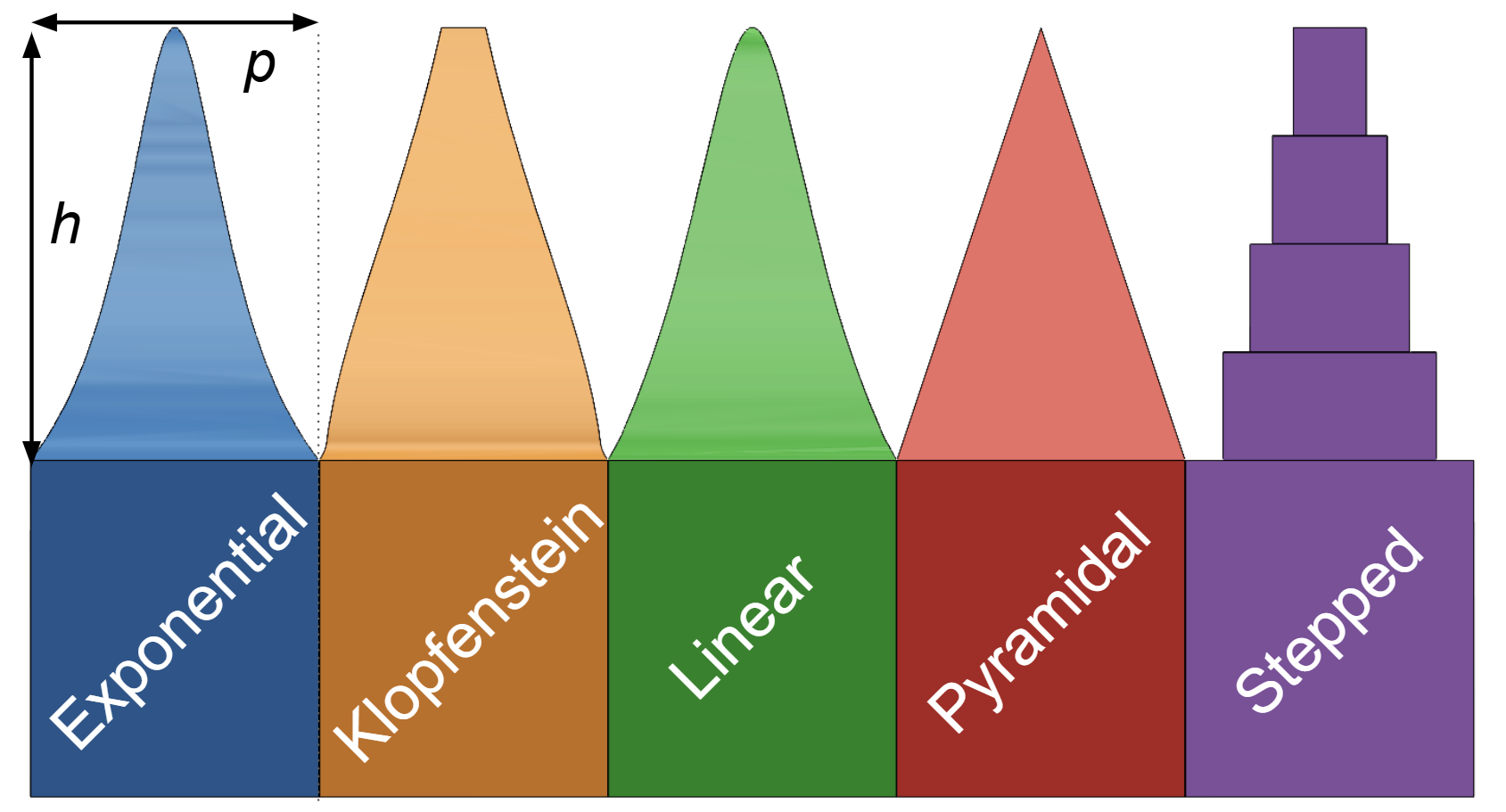}
    \caption{Side view of the different absorber geometries considered in this work along with their names as used in this paper. The aspect ratio of these geometries is $h/p = 3/2$.}
    \label{fig:geometries}
\end{figure}

In this work, we present results from an extensive suite of RF simulations that probe the performance of broadband microwave absorbers as a function of frequency, incidence angles, geometry, and material properties, and compare those results with measurements of various 3D printed absorber candidates. The absorbers that we study are composed of periodic structures with wavelength-scale unit cell sizes, where both geometrical optics and effective medium theory (e.g.\ transfer matrix methods) are insufficient to accurately predict full diffractive properties of these structures \cite{ balanis2012advanced, Taflove2013}. We focus on the 70--\SI{200}{\giga\hertz} frequency range (a free-space wavelength of 4.3-\SI{1.5}{\milli\metre}), as this range overlaps with the 90- and \SIadj{150}{\giga\hertz} frequency bands used by most experiments observing the CMB. The simulation results are validated on specular and non-specular reflectance measurements of 3D printed absorbers with a \SIadj{6}{\milli\metre} unit cell size. 

The work provides clear delineation between specular reflectance and total absorption, highlighting the difference between the two performance metrics. We restricted our study to three microwave-absorber materials characterized solely by a presumed frequency-independent complex dielectric permittivity. These are $\varepsilon _1 = 9.8 + 0.2i$ \cite{otsuka2021material}, $\varepsilon _2 = 3.8 + 1.7i$ \cite{TKRAM}, and $\varepsilon _3 = 5.4 + 0.8i$ \cite{Singh2024}, chosen to match values reported in recent literature. The corresponding loss tangents for these three materials are $\tan \delta_1 = 0.02$, $\tan \delta_2 = 0.45$, and $\tan \delta_3 = 0.15$, respectively. Given that the complex permittivities and loss tangents for these three materials differ quite significantly, we naively expect to see substantial difference in the predicted specular reflectance and total absorption for the three materials. A key goal of this work is to understand to what extent those assumptions hold true. We focus on the $\varepsilon _3$ material in our analysis as we have the capability to 3D print structures with complex permittivities that match this value \cite{Singh2024} (see Section~\ref{sec:design}). 

\section{Absorber design}
\label{sec:design}
\begin{figure}[]
    \centering
    \includegraphics[width=1.0\linewidth]{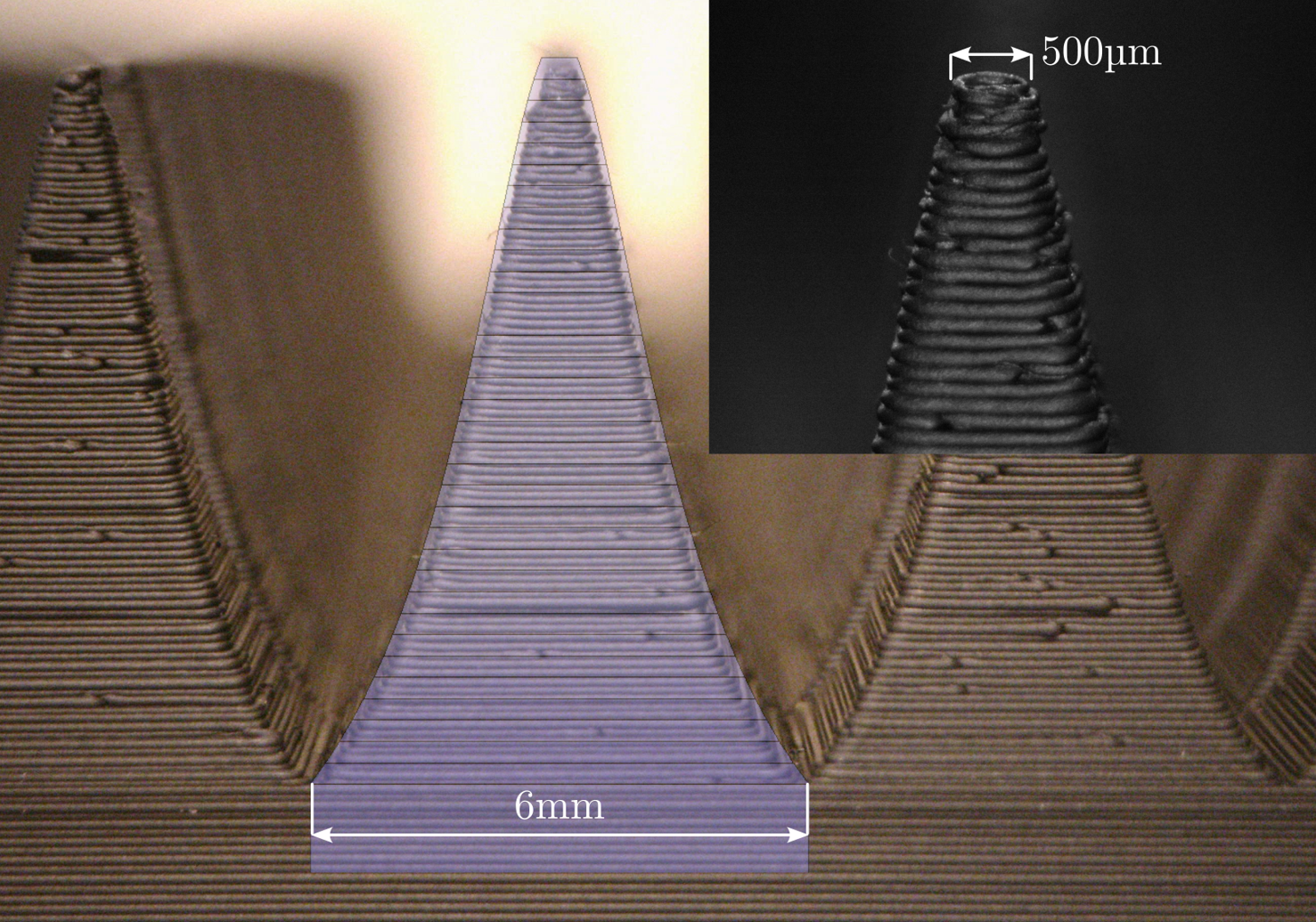}
    \caption{A close up image of a 3D printed HIPS absorber with a linear impedance profile overlaid. The profile is truncated at a tip width of \SI{0.5}{\milli\metre}.}
    \label{fig:cross_section}
\end{figure}

We study five impedance tapers for our absorber candidate (see Figure \ref{fig:geometries}) derived from analytical descriptions of one-dimensional tapers that have been investigated in various publications \cite{Klopfenstein1956, Schroder2015, Pisano2018} (see Appendix A). All of the geometries that we study have a square base of width $p$ and height $h$. As is well known, the amount of reflection scales with the absolute value of the permittivity while the loss in the material depends on the loss tangent (amplitude of complex permittivity). It is desirable to have a high loss term, but this puts stronger constraints on the free-space impedance matching of the absorber surface. 

To study the effect of unit cell size, both $p$ and $h$ were scaled while keeping the aspect ratio constant at $h/p = 1.5$. Larger $h/p$ aspect ratios will improve free-space impedance matching and therefore reduce reflections from these absorbers \cite{pozar2021microwave}. However, real-world applications of broadband absorbers usually restrict the height due to space and manufacturing constraints. Similarly, large aspect ratios put strong demands on material properties due to the risk of the tips of these structures breaking from vibrations or mishandling. We therefore chose to study a number of absorber heights and impedance profiles while keeping the aspect ratio fixed. Since the aspect ratio is fixed, we expect more reflectance in the absorbers candidates with small scale size relative to the wavelength because of the reduced length and volume of the impedance matching geometry. 

The thickness of the base layer is kept constant at \SI{7}{\milli\metre} in all simulations so that the absorption in the bulk is independent of the unit cell. For a base layer of this thickness, at least \SI{99.9}{\percent} of plane wave radiation at \SI{70}{\giga\hertz} is dissipated in the case of the $\varepsilon _2 = 3.8 + 1.7i$ and $\varepsilon _3 = 5.4 + 0.8i$ materials, assuming normal incidence and reflective backing. Significantly thicker base layer is required to produce that much loss for the $\varepsilon _1 = 9.8 + 0.2i$ material due to the lower loss tangent. 

The unit cell size is studied over five octaves, spanning $p = 0.5$, 1.0, 2.0, 4.0, and \SI{8}{\milli\metre}. This corresponds to roughly $0.125$--$2\lambda$ at the low frequency range and $0.25$--$4\lambda$ at the high frequency range. Throughout the text, we use the terms "unit cell size" instead of "scale size" to refer to different values of $p$. However, we note that the term "unit cell size" can be misinterpreted since smaller values of $p$ also imply less absorber material and a more abrubt change in impedance due to the shorter height of these structures.

Given that the refractive index of the materials that are studied ranges from approximately $n \approx 2-3$, the requirement that only zeroth order diffraction modes are allowed to propagate: 
\begin{equation}
p \lesssim \frac{\lambda}{n + \sin \theta _i},
\end{equation}
where $\theta _i$ is the angle of incidence, is only barely met for the smallest unit cell sizes studied in this work \cite{Grann94}. 

We are able to manufacture the five chosen shapes using a Prusa 3D printer with a High Impact Polystyrene (HIPS) filament down to a unit cell size of $p = \SI{6}{\milli\metre}$ (see Figure~\ref{fig:cross_section}), but higher-performing printers and more elaborate manufacturing techniques such as injection molding can easily create structures that contain pyramidal shapes down to \SIadj{0.5}{\milli\metre} unit cell size. Although this work focuses on simulations, comparisons between simulations and measurements are presented in Section~\ref{sec:simvsmeas}.

The HIPS material is produced using a commercial \SIadj{1.75}{\milli\metre} diameter filament that is doped with carbon black of unknown density and particle size \cite{CABELEC}. Our 3D printed samples are measured to have a density of $\rho = \SI{1.05}{\gram \per \cubic \centi \metre}$ and an estimated coefficient of thermal expansion of about \SI{80}{\micro\meter \per \meter \per \kelvin}. Although the low density is great for space-applications with significant mass constraints, the material is expected to have a relatively low thermal conductivity. Measurements of complex permittivity at liquid nitrogen temperatures suggests that the real part of the permittivity reduces by approximately \SI{5}{\percent}. Transmission measurements suggest that the complex permittivity of this material is $\tilde{\varepsilon} _3 = 5.1 + 0.8i$ at 77--$\SI{100}{\kelvin}$.

\section{Simulation setup}
\label{sec:sims}
Using CST (Computer Simulation Technology Studio Suite 2023), we have employed a time-domain solver to estimate the radar cross section of a finite-sized aluminum plate covered with these absorber candidates. Those simulations have been compared with a more common unit-cell simulation approach. The two techniques agree when predicting specular reflectance; however, for the cases with larger unit cell sizes, unit-cell simulation must incorporate a great number of Floquet modes in order to capture the full diffraction pattern made by these structures at these wavelengths. To simplify the analysis, we therefore rely on full-structure time-domain simulations as we are interested in the total absorptance for each absorber.

To compute the reflectance of the absorber, the radar cross section (RCS) is recorded using the CST \textit{Farfield Monitor} and normalized by the RCS of an aluminum mirror. The full-structure simulations capture the response of a $6 \times \SI{6}{\centi\metre\squared}$ absorber plate illuminated by a plane wave. By integrating the reflected power from the absorber and comparing it with the input power, we indirectly measure the amount of power that is absorbed in the material. To verify the validity of this technique, we also run more time-consuming simulations that track the power dissipated in the absorbers and find that the two approaches agree well below the subpercent level. The size of the simulated structure was chosen through a series of convergence tests that indicated that larger structures would not change the predicted reflectance patterns at the level that impacts any conclusions drawn from this work. All simulations are performed on a computing cluster maintained by the Physics Department at Stockholm University. 

\begin{figure}[t!]
    \centering
    \includegraphics[width=1.0\linewidth]{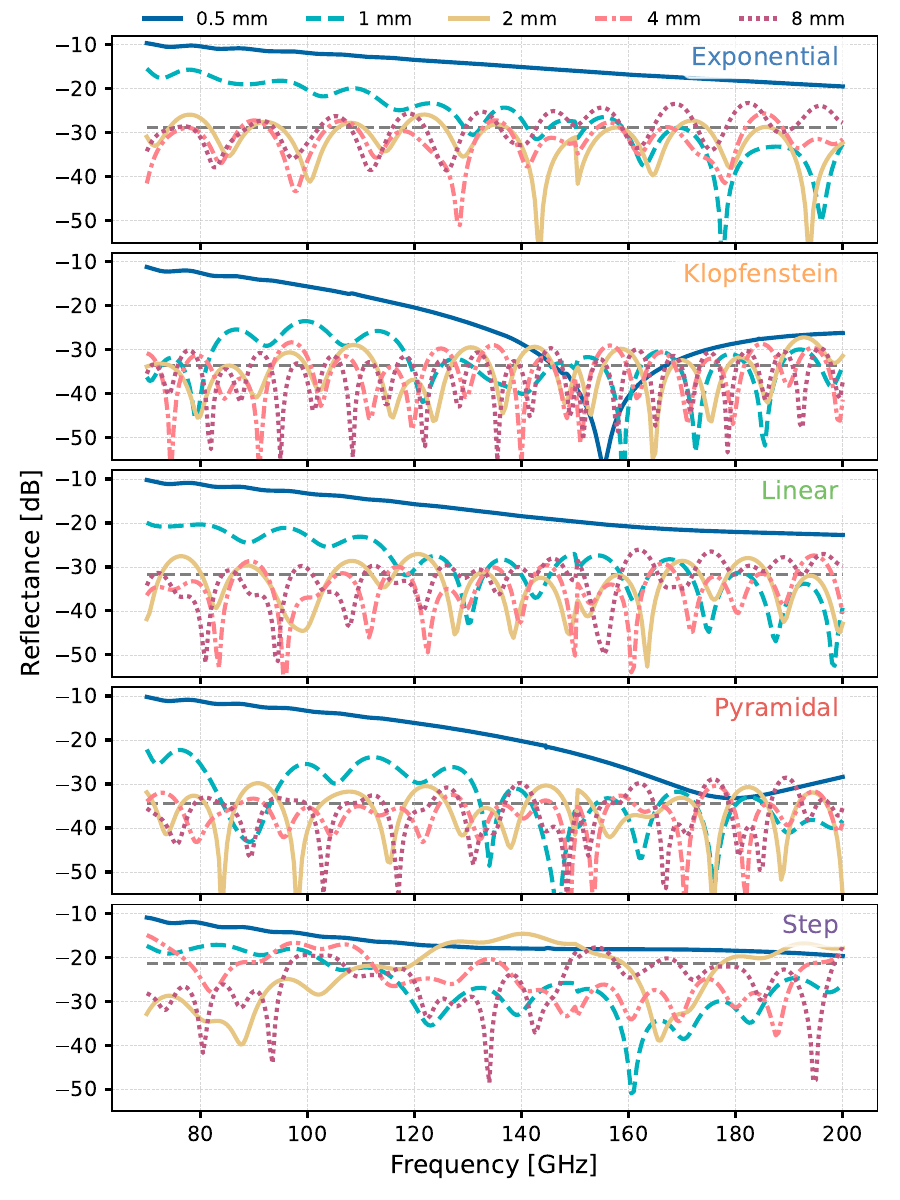}
    \caption{The normalized specular reflectance for on-axis plane-wave illumination as a function of frequency. The five panels show results from 0.5 to \SI{8}{\milli\metre}. The horizontal dashed gray line indicates the combined frequency average for 4 and \SI{8}{\milli\metre} unit cell size. }
    \label{fig:specular}
\end{figure}

\begin{figure}[t!]
    \centering
    \includegraphics[width=1.0\linewidth]{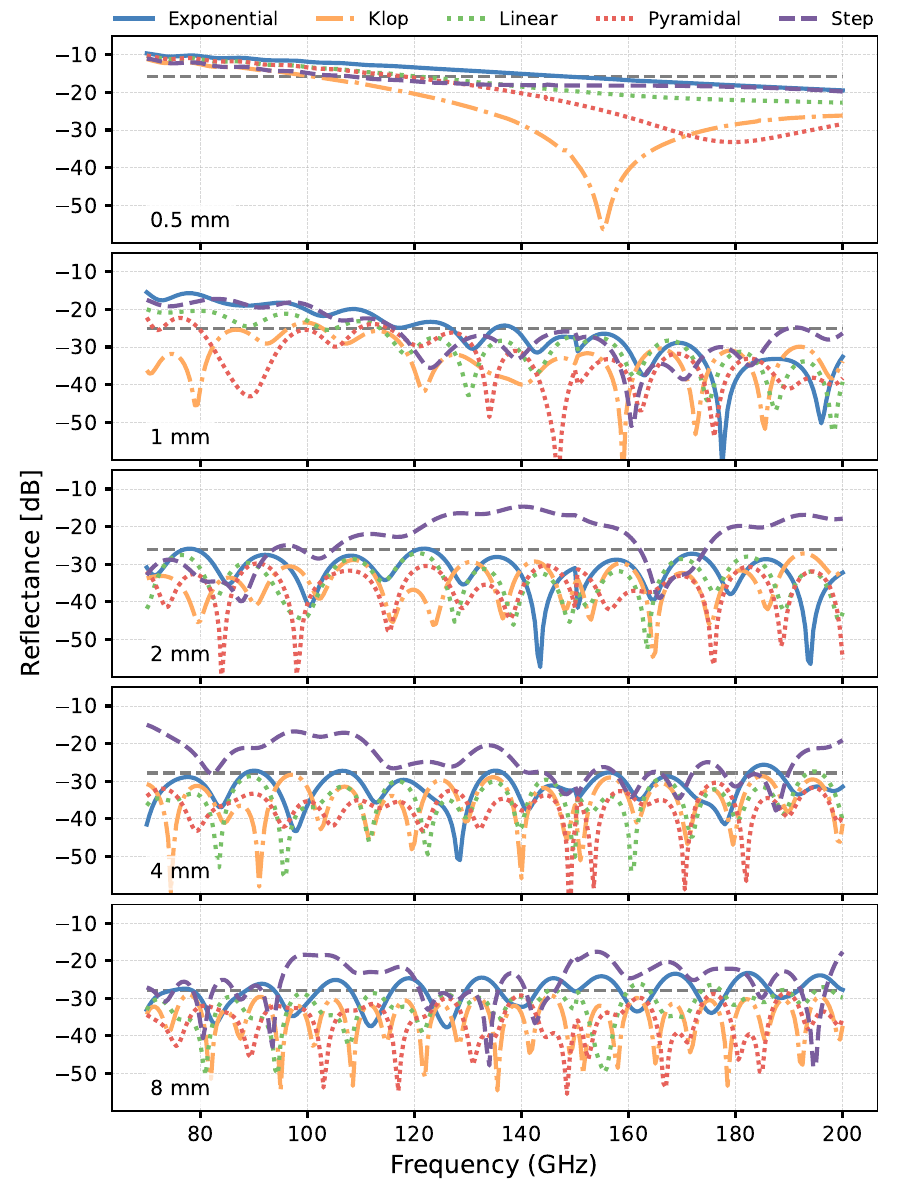}
    \caption{The normalized  specular reflectance for on-axis plane-wave illumination as a function of frequency. The five panels show results for five different unit cell sizes while the five colored lines indicate the different geometries. The horizontal dashed gray line indicates the average specular reflectance for all five geometries across the entire frequency range.}
    \label{fig:specular_unit}
\end{figure}

In these full-structure time-domain simulations, the boundary conditions were set to \textit{Open (add space)} along all three axes. For normal-incidence simulations, magnetic ($H_t=0$) and electric ($E_t=0$) symmetry planes were employed, reducing the modeled volume to ${1}/{4}$ of the original. Simulations were run on a Linux node with \SI{748}{\giga\byte} of RAM and 64 CPU cores with a clock speed of \SI{3.45}{\giga\hertz}. The runtime for each simulation depends on the unit cell size and to a lesser extent the pyramidal profile. A simulation with normal incidence and unit cell size of \SI{1}{\milli\metre} requires around \SI{70}{\giga\byte} of memory to mesh and takes approximately 6-8 hours to complete. Compared to this, the computation time drops by as much as \SI{35}{\percent} for the \SIadj{8}{\milli\metre} unit cell size. We simulate five unit cell sizes for five geometries and three different materials. The total runtime for the work presented here is about 2-3 weeks. Note that simulations for off-axis illumination can easily triple the memory consumption and run time by a factor of 4.

\section{Specular reflectance}

We use the simulations described in Section~\ref{sec:sims} to investigate how specular reflectance at normal angle of incidence depends on frequency, geometry, and unit cell size for the HIPS 3D printing filament ($\epsilon _3$). This corresponds to the property that is most frequently probed with lab measurements. 

To calculate the normalized reflected specular power, we integrate the far field pattern predicted from simulations over a solid angle, $\Omega_{\mathrm{spec}}$, defined by a cone formed by the full-width at half maximum (FWHM) of the reflected radiation pattern: 
\begin{equation}
B_{\mathrm{spec}}(\nu) = \iint_{\Omega_{\mathrm{spec}}} \hspace{-1em} P_\mathrm{samp}(\theta,\phi,\nu) d\Omega  \: \bigg/ \iint_{\Omega_{\mathrm{spec}}} \hspace{-1em} P_\mathrm{Al}(\theta,\phi,\nu) d\Omega,
\label{eq:Rspec}
\end{equation}
where $P_\mathrm{samp}(\theta,\phi,\nu)$ is the far field radiant intensity from the sample as a function of frequency and angle, measured in \SI{}{\watt\per\steradian}. The integral is normalized by an integral over the same solid angle where the integrand has been replaced with the radiation pattern of a bare aluminum plate with the same cross section. Note that all absorber simulations incorporate an aluminum backing plate.

\subsection{Geometrical dependence}
Figure \ref{fig:specular} groups the specular reflectance of normal plane wave incidence as a function of frequency according to the shape of the absorber impedance matching profile of the absorber. It is clear from these simulations that the smaller unit cell-size structures have higher specular reflectance regardless of the impedance taper. This general result might depend on the complex dielectric permittivity, but we note that the same overall result can also be seen in the $\epsilon_1$ and $\epsilon_2$ materials (see also discussion in Section~\ref{sec:total}).

The gray horizontal line indicates the frequency-averaged reflectance for the 4- and \SIadj{8}{\milli\metre} unit cell sizes. This suggests that of the five geometries studied, the Klopfenstein and pyramidal impedance taper demonstrate the lowest specular reflectance across the 70 to \SI{200}{\giga\hertz} frequency range of $-33.5$ and \SI{-34.5}{\deci\bel}, respectively, while the stepped pyramidal geometry appears to perform considerably worse with a specular reflectance of \SI{-19.6}{\deci\bel}. It is surprising that the seemingly small difference in impedance profiles between the so-called linear, exponential, pyramidal, and Klopfenstein profiles can result in more than a factor of 4 difference in specular reflectance averaged over this frequency range. The choice of impedance taper clearly plays an important role in this regard. 

\subsection{Unit cell size dependence}

Figure \ref{fig:specular_unit} shows the same results as in Figure \ref{fig:specular}, but now grouped according to unit cell size. Regardless of the absorber cross section, we see a dramatic change in specular reflectance as we go from subwavelength unit cell sizes to sizes that are comparable or larger than the wavelength. Note that 70 and \SI{200}{\giga\hertz} correspond to approximately 4- and \SIadj{2}{\milli\metre} wavelength in free space, respectively. We see, for example, that the average specular reflectance tends to decrease with increasing unit cell size, but this trend tapers off as the unit cell size becomes comparable to the wavelength. At a unit cell size of \SI{0.5}{\milli\meter}, the average reflectance across the frequency range and the five geometries is only \SI{-17}{\deci\bel}. This number drops down to $-26.4$ and \SI{-27.7}{\deci\bel} for 4- and \SIadj{8}{\milli\metre} unit cell sizes, respectively. Note also that 4 and \SI{8}{\milli\metre} appear to show relatively little difference in specular reflectance for most geometries. 

These results highlight that, while the absorber scale size plays a crucial role in defining specular reflectance, we still see strong dependence on the impedance profile shape. Understanding the interplay between unit cell size and cross sectional geometry is essential for optimizing absorber design when targeting specific frequency bands. 

\subsection{Material dependence}
\begin{figure}[t!]
    \centering
    \includegraphics[width=1.0\linewidth]{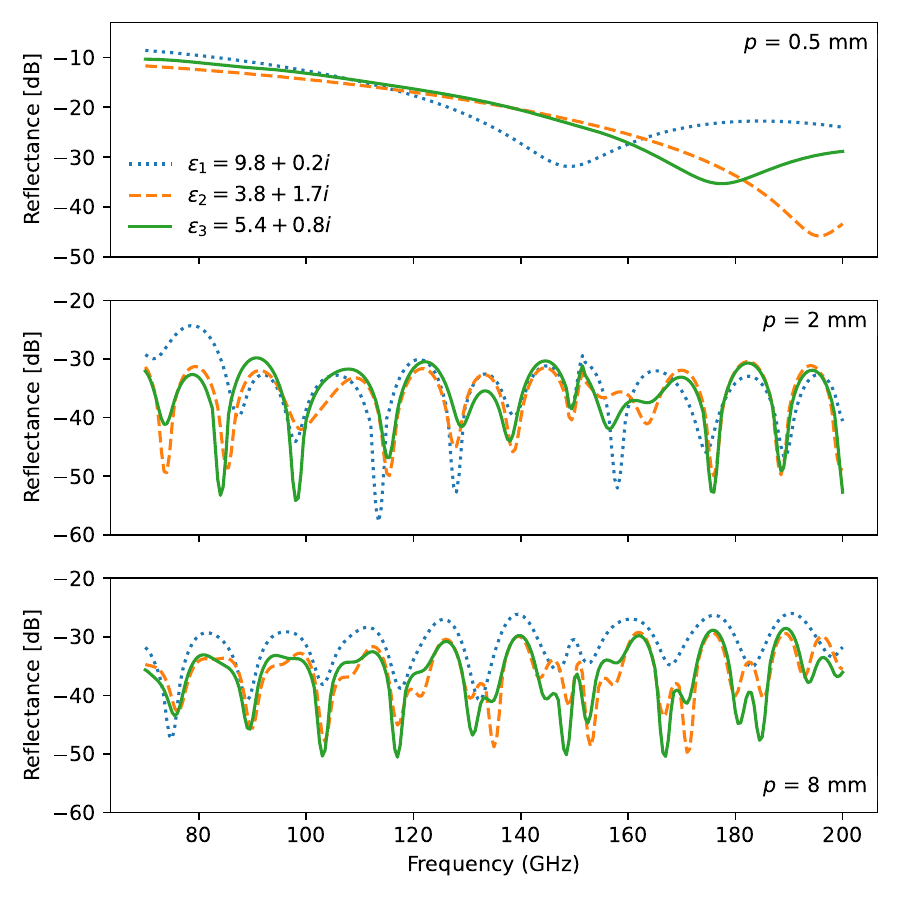}
    \caption{Specular reflectance for the pyramidal structures with three different dielectric materials (\( \varepsilon_1 \), \( \varepsilon_2 \), \( \varepsilon_3 \)) and unit cell sizes of 0.5, 2.0, and \SI{8.0}{\milli\metre}. The discontinuity at \SI{150}{\giga\hertz} is a numerical interpolation artifact due to the splitting of these simulations into two batches, one covering 70-\SI{150}{\giga\hertz} the other covering 150-\SI{200}{\giga\hertz}.}
    \label{fig:Material_properties}
\end{figure}

We have established that unit cell size and impedance tapers crucially impact the specular reflectance of these absorber candidates. Using this simulation pipeline, we can also study the importance of material properties. The specular reflectance for pyramidal geometries as a function of frequency is shown in Figure \ref{fig:Material_properties} with the dotted, dashed, and solid line representing $\varepsilon _1$, $\varepsilon _2$, and $\varepsilon _3$, respectively. Despite significant differences in dielectric properties (see Section~\ref{sec:intro}), the specular reflectance spectra across the three unit cell sizes show surprisingly low variation between materials. This suggests that within the investigated frequency range (70–200~GHz), the specular reflectance is dominated by the geometric configuration and periodicity of the structure, rather than the intrinsic properties of the material. At a unit cell size of $p = \SI{2}{\milli\metre}$, the frequency-average specular reflectance is $-32.5$, $-34.8$, and \SI{-34.4}{\deci\bel} for $\varepsilon_1$, $\varepsilon_2$, and $\varepsilon_3$, respectively, showing that, for this metric, the performance of the three materials is separated by less than a factor of 2. On the other hand, material properties do affect the peak reflectance and resonance as can be clearly seen from the $p = \SI{0.5}{\milli\metre}$ results. The absorber with the highest refractive index, $\varepsilon _1 = 9.8 + 0.2i$, gives the highest specular reflectance for small unit cell sizes as it struggles to match the impedance between free space and the bulk of the material in a short distance.  

\section{Total absorption}
\label{sec:total}

\begin{figure}[t!]
    \centering
    \includegraphics[width=1.0\linewidth]{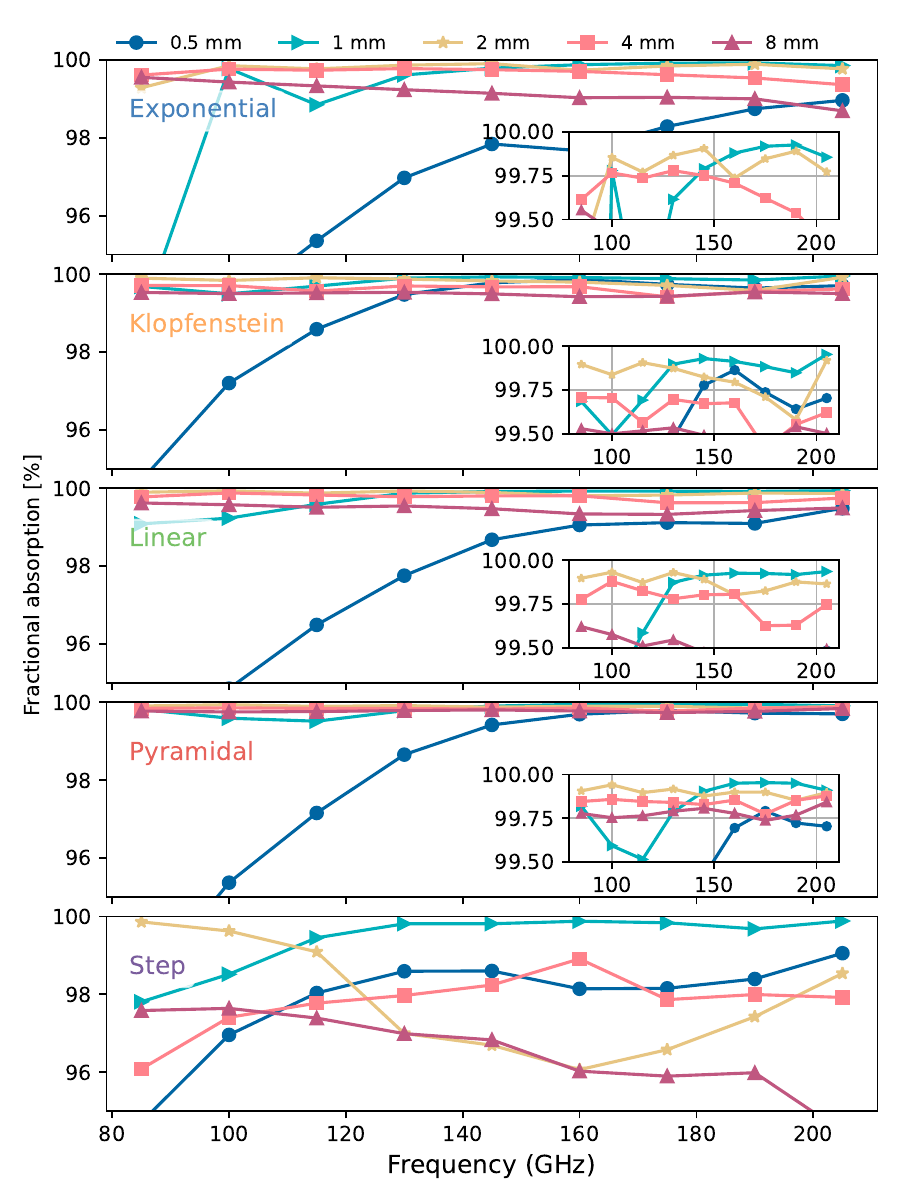}
    \caption{Fractional absorption of the different impedance matching profiles as a function of frequency. Markers show simulated frequencies with lines drawn to connect the markers. The five colors represent the five unit cell sizes studied in this work. Inset axes show a zoomed in version.}
    \label{fig:absorption}
\end{figure}

So far, the results that we have presented only summarize performance for on-axis specular reflectance. However, when illuminated, these absorbers will redistribute power into the half-sphere. We can use the time-domain simulations described in Section~\ref{sec:sims} to probe whether the relatively low specular reflectance that we see for the larger unit cell sizes correlates with total absorption (see Section~\ref{sec:total}).

We track the amount of power that is terminated in the absorber material with the full-structure time-domain simulations. The simulations are performed at \SIadj{7.5}{\giga\hertz} frequency intervals due to the high computation time. This appears to adequately sample the general frequency variation, although we cannot rule out that some resonant behavior is missed. A plane wave hits our absorber candidates at normal incidence, and the power that is not absorbed is reflected or scattered back into the half-sphere. 

To calculate the fraction of power that is reflected or scattered back into the half-sphere, $\Omega _{1/2}$, we integrate the far field pattern predicted from simulations over a $2\pi$ solid angle according to: 
\begin{equation}
R_{\mathrm{scat}}(\nu) = \iint _{\Omega _{1/2}} \hspace{-1em} P_\mathrm{samp}(\theta,\phi,\nu) d\Omega / (I A_\mathrm{plate})
\end{equation}
where $P_\mathrm{samp}(\theta,\phi,\nu)$ is the far field radiant intensity (see Equation~\ref{eq:Rspec}), $I \approx \SI{1.327e-3}{\watt\per\square\meter}$ is the plane wave irradiance that hits the plate and $A$ is the area of the plate. In the limit where all power is reflected back into the half sphere, this expression evaluates to unity. From this, we calculate the fractional absorption, $A(\nu)$, according to $A(\nu) = 1 - R_{\mathrm{scat}}(\nu)$.

Figure~\ref{fig:absorption} shows the fractional absorption grouped by the type of impedance taper. We plot the exponential, Klopfenstein, linear, and pyramidal impedance tapers on the same scale, while the stepped profile shows significantly lower fractional absorption except for one unit cell size (\SI{1}{\milli\meter}). 

Inset figures indicate that the fractional absorption can reach up to \SI{99.9}{\percent} for the 1-, 2-, and \SIadj{4}{\milli\metre} unit cell sizes. At low frequencies (large wavelengths) it is apparent that the $p = \SI{0.5}{\milli\metre}$ absorber candidates have significantly lower fractional absorption. In contrast, the $p = \SI{8}{\milli\metre}$ candidates do not perform as well as the intermediate unit cell sizes for the exponential, Klopfenstein, linear, and pyramidal impedance tapers. This seems to suggest that the best performance is achieved when the unit cell size is comparable to the wavelength. 

Across the full frequency range studied here (70--\SI{200}{\giga\hertz}), the 2- and \SIadj{4}{\milli\metre} unit cell sizes absorb on average 99.71, 99.71, 99.81, and \SI{99.87}{\percent} for exponential, Klopfenstein, linear, and pyramidal absorber profiles, respectively. It is interesting to note that for the impedance profiles that are studied, the traditional pyramidal impedance taper demonstrates the highest fractional absorption for on-axis illumination. 

An extension of this work would include the off-axis illumination of the different impedance profiles for both polarizations. Since broadband microwave absorbers are frequently used in environments where absorption at oblique incidence angles is equally important as normal incidence.

\section{Dependence on material properties}
\begin{figure}[t!]
    \centering
    \includegraphics[width=1.0\linewidth]{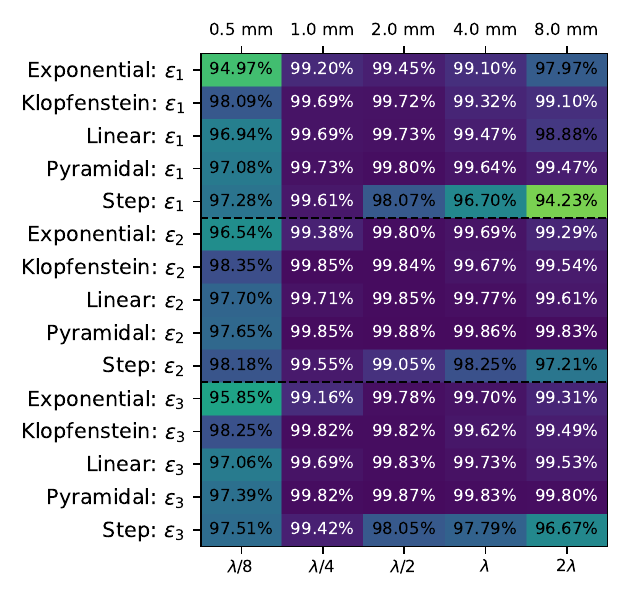}    
    \caption{The average fractional absorption for plane wave at normal incidence integrated across 70 to \SI{200}{\giga\hertz} for three materials ($\varepsilon _1$, $\varepsilon _2$, $\varepsilon _3$), five taper geometries and five unit cell sizes (horizontal axis). The horizontal index of $\lambda /8$ corresponds to $p = \SI{0.5}{\milli\metre}$ and the lower free-space wavelength range of roughly $\lambda = \SI{4}{\milli\metre}$. Similarly, the $\lambda /4$ index corresponds to $p = \SI{1.0}{\milli\metre}$ and $\lambda = \SI{4}{\milli\metre}$, and so forth. Entries that exceed $\SI{99}{\percent}$ are colored in white font, others are in black.}
    \label{fig:table}
\end{figure}

Figure \ref{fig:table} compiles simulation results of fractional absorption for three materials ($\varepsilon _1$, $\varepsilon _2$, $\varepsilon _3$), five geometries, and five unit cell sizes $p = 0.5$, 1.0, 2.0, 4.0, and \SI{8}{\milli\metre}. We express the unit cell sizes using multiples of the the free-space wavelength $\lambda = \SI{4}{\milli\metre}$ which corresponds approximately to the longest wavelength studied in this work. Figure \ref{fig:table} summarizes the results that are displayed in Figure~\ref{fig:absorption}, but now also incorporating simulations for the $\varepsilon _1$ and $\varepsilon _2$ materials. Note that the figure only quantifies fractional absorption for normal incidence.

Out of all of the cases that were considered, we find that the best performance is achieved for a pyramidal taper with a \SIadj{2}{\milli\metre} unit cell size and the $\varepsilon _2$ material. In this case, the simulations predict that on average across the frequency band, $\SI{99.88}{\percent}$ of all radiation will be absorbed. We note however, that there are a range of geometries that perform quite well as long as the unit cell size is either 1 or \SI{2}{\milli\metre}, regardless of the three materials that were considered. This shows that the material property has a much weaker impact on the total absorption than the unit cell size.

Out of the five geometries that we studied, the stepped geometry is found to perform the worst, but the \SI{1}{\milli\metre} unit cell size is clearly an exception. In this case, the stepped geometry performs almost as well as the other four geometries, reaching around $\SI{99.5}{\percent}$ absorption for all three materials that were studied. Again, this behavior is found to be mostly independent of the materials we studied.

\section{Simulation vs Measurement}
\label{sec:simvsmeas}

\begin{figure}[t!]
    \centering
    \includegraphics[width=1.0\linewidth]{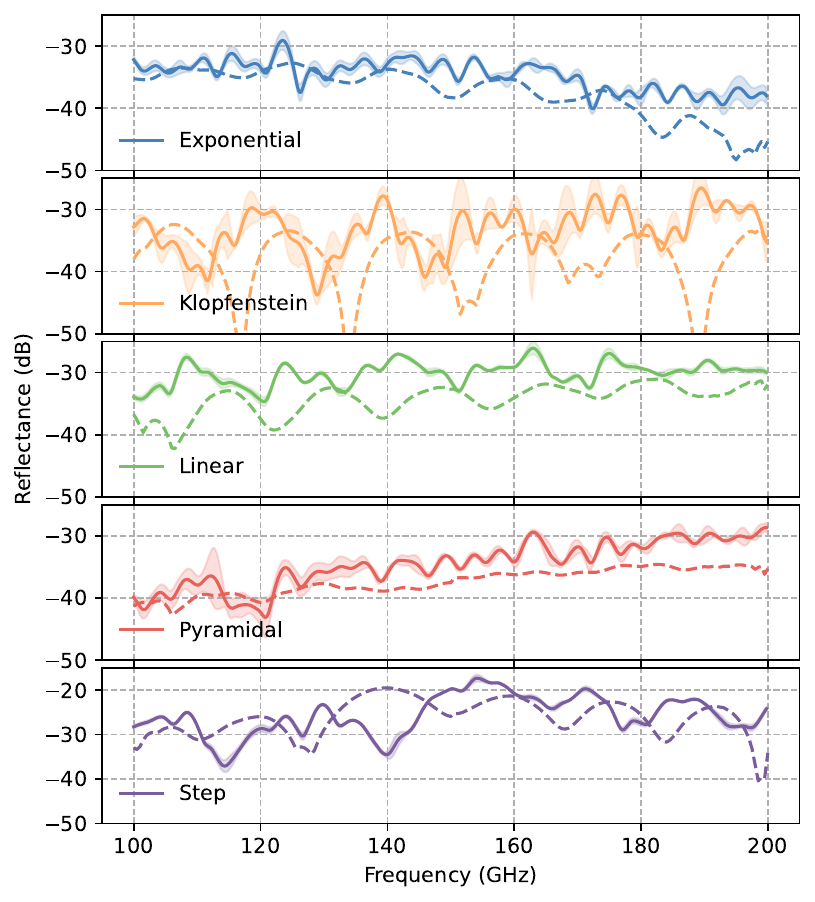}    
    \caption{Comparison of experimental and simulated specular reflectance at \ang{15} angle of incidence for the five absorber geometries across the frequency range of 100–200 GHz. Solid lines represent experimental data while dashed lines indicate simulation results. The filled in band around the solid lines represent the experimental repeatability of these measurements. The vertical axis on the first four panels spans $-25$ to \SI{-50}{\deci\bel} while the panel showing the stepped geometry spans a reflectance of $-15$ to \SI{-50}{\deci\bel}.}
    \label{fig:simvsexperiment}
\end{figure}

To partially validate the simulation results, we have fabricated some of the pyramidal geometries that have been simulated. These geometries were produced using High Impact Polystyrene (HIPS) with a Fused Deposition Modeling (FDM) 3D printing process. We have 3D printed absorber plates with the five geometries studied in this work at a scale of $12 \times \SI{12}{\centi\metre}$, but due to restrictions in the achievable print resolution, we have restricted our analysis to a \SIadj{6}{\milli\metre} unit cell. 

Using a TOPTICA measurement system, we have measured the complex permittivity of the HIPS material and found that it can be approximated as $\varepsilon _3 = 5.4 + 0.8i$ across the entire frequency range of interest \cite{Singh2024}. Using that same system, we also measured the specular and nonspecular reflectance for each of the five samples keeping the incidence angle fixed at \SI{15}{\degree} with the angle of the receiver varied from \SI{15}{\degree} to \SI{45}{\degree} in increments of \SI{7.5}{\degree}. Figure~\ref{fig:simvsexperiment} shows measurement results (solid lines) for specular reflectance of s-polarization (TE) along with the simulated response (dashed lines). Each measurement is repeated four times and the sample is repositioned in the sample holder between measurements to quantify the repeatability of our measurements. The colored band around the solid lines indicates the scatter between each of those four measurements. We note that the variation is particularly strong for the Klopfenstein geometry, potentially due to manufacturing non-idealities. 

On average across the 100-\SI{200}{\giga\hertz} frequency band, the measurements show a factor of 1.2--2.5 higher specular reflectance compared to simulation predictions, depending on the impedance taper. Despite systematically measuring higher reflectance, these results closely follow the simulation trends, especially the variation with frequency. Differences between measurements and simulations at this magnitude can likely be attributed to non-idealities in the 3D printed shapes and the alignment of the measurement system. Note that spillover from transmitter to receiver at the level of \SI{0.05}{\percent} compared to the total transmitter power can explain differences at the  \SI{-33}{\deci\bel} level.

The substantial frequency variations with periods of 10-\SI{20}{\giga\hertz} seen in both simulations and measurements for the Klopfenstein, linear, and stepped geometries suggest that the measurements are capturing expected geometrical effects quite well. Note, for example, that both simulations and measurements suggest a slight drop in reflectance with frequency for the exponential geometry, while converse is seen in the pyramidal geometry. Of the five geometries that we studied, the exponential taper shows the lowest frequency-averaged specular reflectance of \SI{-34.1}{\deci\bel} across 100-\SI{200}{\giga\hertz}. The pyramidal and Klopfenstein taper follow closely with a reflectance of $-33.3$ and \SI{-32.2}{\deci\bel}, respectively.

\section{Conclusions}
We conducted a parametric study of square-tiled broadband absorbers operating over the 70–\SI{200}{\giga\hertz} frequency range. Assuming a fixed aspect ratio, we find that the critical design parameter governing both specular reflectance and total absorption is the absorber unit cell size. The best performance occurs when the unit cell dimension is comparable to the free-space wavelength—approximately 1–\SIadj{4}{\milli\metre} in our study. Within this range, the exponential, Klopfenstein, linear, and pyramidal tapers all achieve more than \SI{99.7}{\percent} average absorption across the band, with the traditional pyramidal taper performing best overall.

At the upper end of the frequency range ($\nu>\SI{150}{\giga\hertz}$), the \SIadj{1}{\milli\metre} unit cell generally yields the highest fractional absorptance, largely independent of taper profile. For the parameters explored, the influence of material properties is comparatively weak; the geometry and unit cell size dominate the absorber performance.

When considering specular reflectance in isolation, the effect of taper shape is most pronounced for sub-wavelength unit cells and becomes negligible as the cell size approaches or exceeds the free-space wavelength. The stepped taper stands out as a clear outlier, exhibiting substantially poorer absorption across nearly all unit cell sizes. Some optimization of the stepped geometry could potentially improve its performance as an absorber.

If one were to rely solely on specular reflectance as a performance metric, it might appear that absorbers with 4- or \SIadj{8}{\milli\metre} unit cells outperform those with 0.5-, 1-, or \SIadj{2}{\milli\metre} cells (see Figures~\ref{fig:specular} and \ref{fig:specular_unit}). However, the frequency-averaged fractional absorption summarized in Figures~\ref{fig:absorption} and \ref{fig:table} indicates that 1–\SIadj{2}{\milli\metre} unit cells absorb the largest fraction of incident radiation in the 70–\SI{200}{\giga\hertz} range. Since broadband microwave absorber performance is largely determined by fractional absorption rather than fractional reflectance, the latter should not be used in isolation to assess absorber efficacy. Nevertheless, specular reflectance remains both easier to model and measure, and thus is often used as a proxy in the literature.

To partially validate our simulations, we fabricated \SIadj{6}{\milli\metre}-pitch HIPS samples and measured their specular reflectance at \ang{15}. The measured spectra reproduce the simulated frequency trends and relative ranking of taper profiles, with modest gain offsets likely attributable to manufacturing tolerances and small systematic errors in the measurement setup.

\begin{backmatter}
\bmsection{Funding} GS and JEG acknowledges support from the Swedish National Space Agency (SNSA/Rymdstyrelsen) and the Icelandic Research Fund (Grant number: 2410656-051). Funded in part by the European Union (ERC, CMBeam, 101040169). JEG gratefully acknowledges support from the University of Iceland Research Fund. 

\bmsection{Acknowledgment} We thank Ashali Ásrún Gunnarsdóttir for performing material property measurements that validated statements made in this paper.

\bmsection{Disclosures} The authors declare no conflicts of interest.

\end{backmatter}

\bibliography{paper}

\section{Appendix}
\subsection{Impedance tapers}

An effective medium description of absorber impedance tapers is not expected to be fully applicable at scales comparable to the wavelength \cite{Lalanne1996}, since at these scales the local fields of a given unit cell can no longer be simply averaged over and the quasi-static approximation no longer applies. One can still use open-source codes for one-dimensional stratified media to gain some insight into the performance of absorber impedance tapers as a function of geometry and material properties \cite{tmm_github_3, Hileman2013, Luce2022}. Not being able to rely on fast numerical tools to predict the performance of different impedance profiles for absorbers, we instead use simple concepts from impedance theory as a basis for our proposed geometries which we then probe with finite element time-domain simulations.

The impedance of a lossless and non-magnetic ($\mu _r$ = 1) dielectric with relative permittivity $\epsilon _r$ is defined as
\begin{equation}
Z = \sqrt{\frac{\mu _0}{\epsilon_r \epsilon_0}} = \frac{Z_0}{\sqrt{\epsilon_r}}, 
\end{equation}
where $\epsilon_0$ and $\mu_0$ are the electrical permittivity and magnetic permeability of free space, respectively. The free-space impedance is defined as $Z_0 \equiv \sqrt{ \mu_0  / \epsilon_0}$. 

The electrical permittivity of materials considered for microwave absorbers differs significantly from that of free space, which leads to reflections at interfaces. The performance of absorbers with anisotropic geometries that need to work for a wide range of frequencies and incidence angles is therefore critically dependent on the details of the geometry, including the length scales of those geometries relative to the wavelength. Several geometries are considered in the literature \cite{wollack2014cryogenic, pozar2021microwave, Klopfenstein1956, petroff20193d, Han2016}.

For the absorbers discussed here, an impedance gradient is realized by an array of pyramid-like structures with square unit cells. They are tiled in a square-periodic pattern on the top side of a flat absorber plate which is backed by an aluminum plate (a decent approximation for perfect reflector at microwave frequencies). The width of a single pyramid changes from the thickness of the unit cell at the bottom to a relatively sharp point at the top, where $l=h$. 

\begin{figure}
    \centering
    \includegraphics[width=1.0\linewidth]{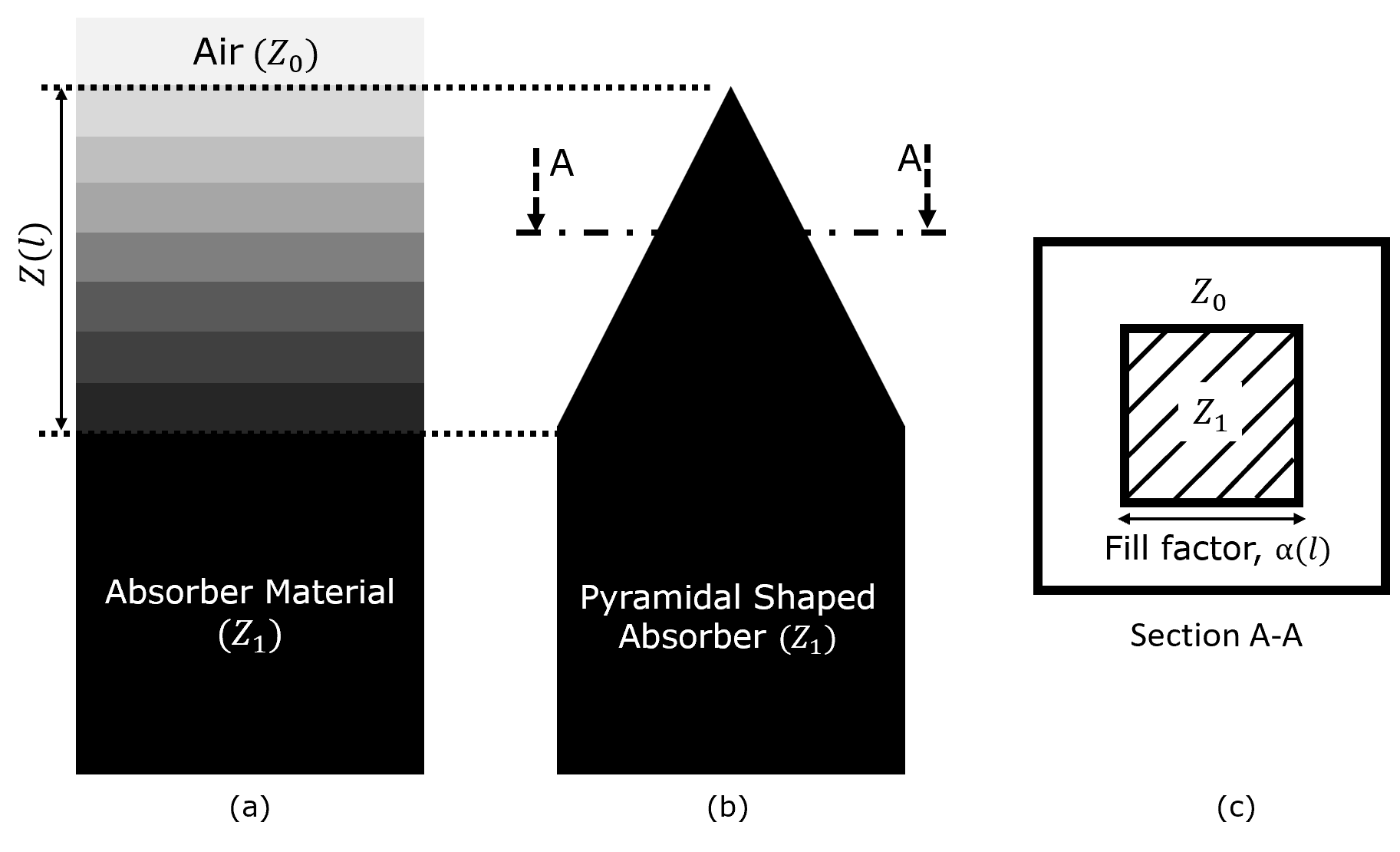}
    \caption{(a) An effective medium representation of the anti-reflection coating (ARC) layer, which transitions from free space to the absorber bulk material. (b) Schematic of a pyramidal-based absorber, designed to gradually transition impedance and reduce reflections. (c) Cross-sectional view at an arbitrary location within the pyramid, showing the fractional composition of the absorber material within a given unit cell.}
    \label{fig:pyramid_schematic}
\end{figure}

The microwave impedance at height $l$ can be estimated as the effective impedance of a thin layer at that height, partially filled with absorptive material and partially with free space (see Figure~\ref{fig:pyramid_schematic}). We can approximate this layer as an isotropic layer with a permittivity calculated as the area-weighted mean. The impedance taper for a pyramid based on multiple layers with different effective permittivities is then reduced to a 1D problem. In this scenario, the area fill factor is position-dependent and we can define an effective impedance, $Z(l)$, that varies with $l$ extending from a location $l=0$ to $l=h$, where $h$ defines the height of the taper. At $l=0$, the impedance matches the impedance of the lossy dielectric, which we define as $Z_1 \equiv Z_0 / \sqrt{\epsilon _r}$,  while $Z(h) = Z_0$ just outside the tip of the taper. With a fixed absorber height, an optimization of the shape of the taper to maximize total absorptance across both polarizations and a range of frequencies and incidence angles can begin. 

Within the taper, we can estimate the impedance using an effective medium approximation. The relative permittivity at location $l$ is calculated as:
\begin{align}
    \varepsilon (l) &= (1 - \alpha (l) ^2) + \alpha (l) ^2 \varepsilon_r, \nonumber \\
                    &= 1 + ( \varepsilon_r-1)\alpha (l) ^2,
\label{eq:el}
\end{align}
where $\alpha (l)$ is the unit cell fill factor as a function of location $l$ (see Figure~\ref{fig:pyramid_schematic}). With the effective permittivity known, the impedance is then calculated as 
\begin{equation}
Z(l) = Z_0/\sqrt{\varepsilon (l)}.
\label{eq:Zleps}
\end{equation}
For any desired impedance profile $Z(l)$, the unit cell fill factor $\alpha (l)$, can also be calculated easily according to 
\begin{equation}
\varepsilon (l) =  (Z_0 / Z(l))^2,
\end{equation}
which through the use of Equation~\ref{eq:el} gives 
\begin{equation}
\alpha(l) = \sqrt{\frac{(Z_0 / Z(l))^2 - 1}{\varepsilon_r-1}}.
\label{eq:alphal}
\end{equation}
The dielectrics that we consider in this paper are lossy and therefore modeled using a permittivity with a non-zero complex component. However, when calculating the impedance profiles, we only consider the real component of the permittivity. We assume $Z_1 = Z_0 / \sqrt{5.4}$ where $5.4$ corresponds to the real part of the complex permittivity $\varepsilon _3 = 5.4 + 0.8i$. The following sections provide the functional description of the impedance profiles studied in this paper. 

\subsubsection{Pyramidal Taper}
The simplest impedance profile is probably the pyramidal taper which is shown in Figure~\ref{fig:pyramid_schematic}. In cross section, simple pyramidal absorbers are drawn by extending a line from the base of the absorber unit cell to the tip of the absorber making some angle $\theta = \arctan (p/2h)$ relative to the surface normal. The area fill fraction of the absorber will scale like
\begin{equation}
\alpha(l) = (p-2\tan(\theta)l)^2 / p^2,
\end{equation}
 where $p$ is the linear dimension of the unit cell size. Unlike the following geometries, this periodic structure is not defined by specifying $Z(l)$, but rather by the real-space shape of the structure. The corresponding effective impedance can be calculated using Equations~\ref{eq:el} and \ref{eq:Zleps}. Pyramids are widely used as anti-reflection coatings on various kinds of absorbers~\cite{chesmore2021simons, kubytskyi2012fast, Pisano2018}. 

\begin{figure}[t!]
    \centering
    \includegraphics[width=1.0\linewidth]{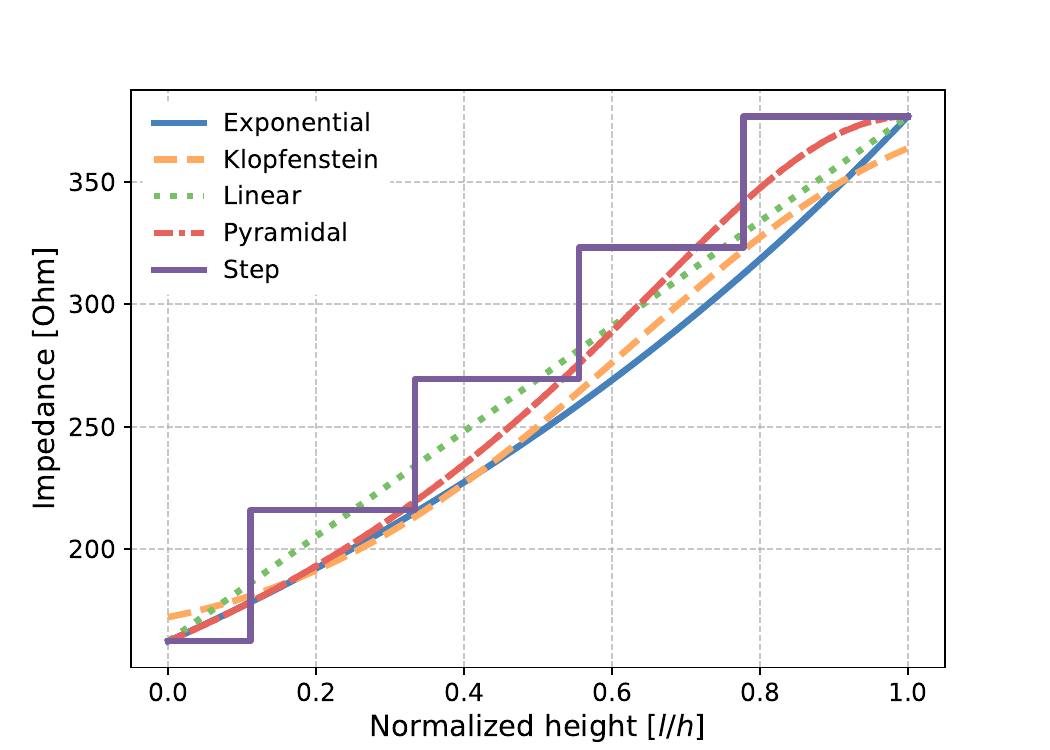}
    \caption{Variation of the different impendance profile along the length of the pyramid from the base $(Z_1)$ to the top  $(Z_0)$.}
    \label{fig:Impedance_comparison}
\end{figure}

Figure~\ref{fig:Impedance_comparison} shows the variation of the impedance from the bottom of the pyramid to the tip for the different impedance profiles. Note that these profiles are normalized to the height of the impedance taper. Note that absorbers with small scale size, e.g., $p=\SI{0.5}{\milli\meter}$ which implies $h=\SI{0.75}{\milli\meter}$ for a $3/2$ aspect ratio, will have a more abrupt transition in impedance between $Z_0$ and $Z_1$ compared to absorbers where $p = \SI{8}{\milli\meter}$.

\subsubsection{Exponential Taper}
For a rising taper (e.g., from absorber to air; $Z_1<Z_0$), two arbitrary points with different impedances can be joined by an exponential curve as follows \cite{pozar2021microwave}:
\begin{equation}
    Z_{\text{exp}}(l)  = Z_1 \exp\left[\frac{l}{h} \ln \left( \frac{Z_0}{Z_1}\right)\right],
\label{eq:expo_taper}
\end{equation}
where $h$ is the height of the absorber. 

\subsubsection{Klopfenstein Taper}
Grann et al.\ \cite{grann1995optimal} and Kubytskyi et al.\ \cite{kubytskyi2012fast} define the Klopfenstein taper $Z_\mathrm{Klop}(l)$ in terms of refractive index $n(l)$. Here we rewrite it in terms of impedance, $Z(l)$, which agrees with the description of the Klopfenstein taper given by Pozar \cite{pozar2021microwave}. The Klopfenstein taper will not achieve zero reflectance at any frequency. Instead, it shows a ripple in the reflected spectrum~\cite{grann1995optimal, pozar2021microwave}. The magnitude of this ripple is proportional to the parameter $\Gamma_m$ and this parameter is chosen as part of the taper design process. The choice is a trade-off between the ripple magnitude and overall reflectivity magnitude. Here, we chose $\Gamma_m = 0.035$ which agrees with values in the literature \cite{pozar2021microwave, grann1995optimal}. Klopfenstein tapers are found to minimize reflectance when integrated over a finite frequency range, compared to other impedance tapers \cite{pozar2021microwave}. They are also found to achieve the lowest values of reflectance for a fixed height $h$~\cite{pozar2021microwave}. We use the following equations to calculate $Z_\mathrm{Klop}(l)$:

\begin{equation}
    Z_\mathrm{Klop}(l) = \sqrt{Z_0 Z_1} \exp \left[ \Gamma_m A^2 \phi \left( 2 \frac{l}{h} - 1, A \right) \right],
\end{equation}
where the function $\phi(x,A)$ is defined as:
\begin{equation}
    \phi(x, A) = \int_{0}^{x} \frac{I_1 \left( A \sqrt{1 - y^2} \right)}{A \sqrt{1 - y^2}} \, dy,
\end{equation}
and $I_1$ is the modified Bessel function and $A$ is defined as:
\begin{equation}
    A = \cosh^{-1} \left[ \frac{1}{2 \Gamma_m} \ln \frac{Z_1}{Z_0} \right].
\end{equation}

As can be seen in Figure~\ref{fig:Impedance_comparison}, there is a discontinuity at the top and bottom for $Z_\mathrm{Klop}(l)$. This is consistent with \cite{pozar2021microwave}. This behaviour can be controlled by varying $\Gamma_m$; lower the value of $\Gamma_m$, smoother will be the transition at the extreme ends of the impedance profile; however, this also leads to a sharp change in the impedance around the center of the pyramid length. 

\subsubsection{Linear Taper}
A linear impedance taper, $Z_\mathrm{Lin}(l)$, is a profile where the variation from ${Z_0}$ to ${Z_1}$ is linear in $l$. It can be defined as follows \cite{sangam2018linear}:

\begin{equation}
    Z_\mathrm{lin}(l)= \left( \frac{Z_1 - Z_0}{h} \right) \cdot l + Z_0,
\end{equation}
where $h$ is the height of the absorber. The corresponding unit cell factor fill factor can be calculated from plugging this into Equation~\ref{eq:alphal}.

\subsubsection{Step Taper}
The stepped taper investigated here consists of a finite number of equally spaced steps along the vertical direction, sampled using a linear extrapolation between the two extreme values. The number of interfaces between layers was arbitrarily chosen to be $N=5$ which results in 4 intermediate layers with a constant impedance between $Z_0$ and $Z_1$. At each step there is a jump of approximately \SI{43}{\ohm} in impedance. A similar impedance taper is discussed in Refs.~\cite{Pisano2018}. Increasing the number of steps will smoothen the impedance profile and likely reduce reflections. A general equation for the stepped pyramidal taper ($Z_\mathrm{step}(l)$) is as follow:
\begin{equation}
    Z_{\mathrm{step}}(l)= Z_0 + \frac{m}{N}\,(Z_1 - Z_0),
\qquad \Delta l=\frac{h}{N-1},
\end{equation}

where $m\in\{1,\dots,N-1\}$  and $\ (m-1)\,\Delta l < l \le m\,\Delta l$ .

\end{document}